\documentclass[twocolumn,showpacs,prb,altafilletter,runinadress,onlinecite]{revtex4}

\usepackage{graphicx,psfrag,epsf,epsfig}
\usepackage{dcolumn}
\usepackage{bm,bbm}
\usepackage{pstricks,pst-plot}     
\usepackage{latexsym}
\usepackage{mathrsfs,amsmath,amssymb,textcomp} 

\usepackage{hyperref}

\begin{document}

\title{Microscopic theory of absorption and emission in nanostructured solar
cells: \\Beyond the generalized Planck formula}

\author{U. Aeberhard}
\email{u.aeberhard@fz-juelich.de}
\author{U. Rau}

\affiliation{IEF-5: Photovoltaik, Forschungszentrum J\"ulich, D-52425 J\"ulich,
Germany}

\date{\today}

\begin{abstract}
Absorption and emission in inorganic bipolar solar cells based on low dimensional structures exhibiting the 
effects of quantum confinement is investigated in the framework of a comprehensive microscopic theory of the 
optical and electronic degrees of freedom of the photovoltaic system. In a quantum-statistical treatment based 
on non-equilibrium Green's functions, the optical transition rates are related to the conservation of electronic
currents, providing a quantum version of the balance equations describing the operation of a photovoltaic device. 
The generalized Planck law used for the determination of emission from an excited semiconductor in 
quasi-equilibrium is replaced by an expression of extended validity, where no assumptions on the distribution 
of electrons and photons are made. The theory is illustrated by the numerical simulation of single quantum well 
diodes at the radiative limit.
\end{abstract}

\maketitle

\section{Introduction} 

Knowledge about the luminescence of solar cells is of great value both for the characterization of the cell 
and for the analysis of the device performance due to its correspondence to the radiative losses. Established 
theories predicting the luminescent emission of bulk solar cells are mainly based on the generalized Planck  
and Kirchhoff formulas [\onlinecite{wuerfel:82}] relating the emitted photon
flux to the absorptivity and the bias voltage. However, the generalized Planck
spectrum was derived based on the crucial assumption that the carriers are well described 
by a local equilibrium distribution characterized by a single quasi-Fermi level and at the temperature of the 
lattice. This assumption is no longer justified in nanostructured solar cells based on quantum wells or quantum 
dots
[\onlinecite{tsui:96,nelson:97,bremner:99,barnham:00,kluftinger:00,kluftinger:01,honsberg:02,mazzer:06,fuehrer:06,fuehrer:07}].
The purpose of the present work thus is to provide a more general approach to absorption and emission in bipolar semiconductor 
nanostructures where the critical assumptions are relaxed. To meet the requirements formulated above, a description of emission is derived 
within a general microscopic  
nonequilibrium theory of quantum-confined photovoltaic systems
[\onlinecite{ae:prb_08}]. The appeal of the approach lies in the way the optical
properties of a semiconductor nanostructure are related to its electronic properties on very general grounds. 
Based on non-equilibrium quantum-statistical mechanics and starting from a quantum theory of 
electrons and photons in semiconductors nanostructures, absorption and emission are calculated from the 
transverse polarization function, which appears as the convolution of the non-equilibrium Green's functions 
for electrons and holes
[\onlinecite{kadanoff:62,keldysh:65,henneberger:96,pereira:96,pereira:98,faleev:02,shtinkov:04,richter:08,henneberger:09}].
Unlike more basic approaches, the theory allows the inclusion of many-particle corrections and interaction effects, such as those due to 
electron-phonon coupling or exciton formation.

\section{Theory}

The following outline of the general theory of quantum photovoltaic devices is restricted to 
the radiative limit, i.e. non-radiative recombination (Shockley-Read-Hall, Auger) and generation 
(impact ionization) processes are not considered.

\subsection{Hamiltonian}
At the radiative limit, the full quantum photovoltaic device  is described in terms of the model Hamiltonian
\begin{align}
 \mathcal{H}=&\mathcal{H}_{e}+\mathcal{H}_{\gamma}+\mathcal{H}_{p},&\mathrm{total}\qquad \\ 
\mathcal{H}_{e}=&\mathcal{H}_{e}^{0}+\mathcal{H}_{e}^{i},
&\mathrm{electronic}\qquad \\ 
\mathcal{H}_{e}^{i}=&\mathcal{H}_{e\gamma}+\mathcal{H}_{ep}+\mathcal{H}_{ee},
&\mathrm{interaction}\qquad
\end{align}
consisting of the coupled systems of electrons ($\mathcal{H}_{e}$), photons
($\mathcal{H}_{\gamma}$) and phonons ($\mathcal{H}_{p}$). Since the focus is on
the electronic device characteristics, only $\mathcal{H}_{e}$ is considered here,
however including all of the terms corresponding to coupling to the bosonic systems. Within the electronic
part, $\mathcal{H}_{e}^{0}$ contains the kinetic energy, the (bulk) band
structure and band offsets, and it also includes the electrostatic potential from the solution of Poisson's equation. 
The interaction part $\mathcal{H}_{e}^{i}$ consists of the terms
$\mathcal{H}_{e\gamma}$, $\mathcal{H}_{ep}$ and $\mathcal{H}_{ee}$ for the
interactions of electrons with photons, phonons and electrons or holes, respectively.
While the first term describes generation and recombination of free 
charge carriers via the absorption and emission 
of photons, the second one provides the carrier relaxation via the absorption and emission of 
phonons. The last term contains the carrier-carrier interactions responsible for screening effects 
or exciton formation.

Within the non-equilibrium Green's function theory of quantum optics and transport in excited 
semiconductor nanostructures, physical quantities are expressed in terms of quantum statistical 
ensemble averages of single particle operators, namely the fermion field
operator $\hat{\Psi}$ for the charge carriers, the quantized photon field vector
potential $\hat{\mathbf{A}}$ for the photons and the ionic displacement field
$\hat{U}$ for the phonons. The corresponding Green's functions are
\begin{align}
 G(\underbar{1}\underbar{2})=&-\frac{i}{\hbar}\langle\hat{\Psi}(\underbar{1})
 \hat{\Psi}^{\dagger} (\underbar{2})\rangle_{C},\qquad&\mathrm{electrons}\\ 
 D_{ik}^{\gamma}(\underbar{1}\underbar{2})=&-\frac{i}{\hbar}\langle
 \hat{A}_{i}(\underbar{1})\hat{A}_{k}(\underbar{2})\rangle_{C},\qquad&\mathrm{photons}\\
  D_{j}^{p}(\underbar{1}\underbar{2})=&-\frac{i}{\hbar}\langle\hat{U}_{j}^{\dagger} 
  (\underbar{1})\hat{U}_{j}(\underbar{2})\rangle_{C}\qquad&\mathrm{phonons}
 \end{align}   
where $\langle...\rangle_{C}$ denotes the contour ordered operator average 
peculiar to non-equilibrium quantum statistical mechanics
[\onlinecite{kadanoff:62,keldysh:65}] for arguments
$\underbar{1}=(\mathbf{r}_{1},t_{1})$ with temporal components on the Keldysh
contour [\onlinecite{keldysh:65}].
   
\subsection{Dyson's equations}
The Green's functions follow as the solutions to corresponding Dyson's
equations [\onlinecite{pereira:96,pereira:98,schaefer:02}],
\begin{align}
\int d\underbar{3}[G_{0}^{-1}(\underbar{1}\underbar{3})-\Sigma(\underbar{1}\underbar{3})] G(\underbar{3}
\underbar{2})
&=\delta(\underbar{1}\underbar{2}),\\
\int d\underbar{3} [(\overleftrightarrow{D}_{0}^{\gamma})^{-1}(\underbar{1}\underbar{3})-
\overleftrightarrow{\Pi}^{\gamma}(\underbar{1}\underbar{3})]
 \overleftrightarrow{D}^{\gamma}(\underbar{3}\underbar{2})&=
 \overleftrightarrow{\delta}(\underbar{1}\underbar{2}),\\
\int d\underbar{3} [(D_{j,0}^{p})^{-1}(\underbar{1},\underbar{3})-
\Pi_{j}^{p}(\underbar{1}\underbar{3})] D_{j}^{p}(\underbar{3}\underbar{2})&=
 \delta(\underbar{1}\underbar{2}).
 \end{align}
$G_{0}$, $D_{0}^{\gamma}$ and $D_{0}^{p}$ are the propagators for noninteracting
electrons, photons and phonons, respectively,  $\leftrightarrow$ denotes
tensorial transverse quantities and $j$ labels the phonon mode. The electronic
self-energy $\Sigma$ encodes the renormalization of the charge carrier Green's
functions due to the interactions with photons and phonons, i.e. generation, recombination and relaxation processes. Charge injection and absorption at 
contacts is considered via an additional boundary self-energy term reflecting the openness of 
the system. The photon and phonon self-energies $\Pi^{\gamma}$ and
$\Pi^{p}$ describe the renormalization of the optical and vibrational
modes, leading to phenomena such as photon recycling or the phonon bottleneck 
responsible for hot carrier effects.

\subsection{Microscopic optoelectronic conservation laws}
The macroscopic equation of motion for a photovoltaic system is the continuity equation for charge carriers, 	
\begin{align}
\partial_{t}\rho_{c}(\mathbf{r})+
\nabla\cdot
\mathbf{j}_{c}(\mathbf{r})=\mathcal{G}_{c}(\mathbf{r})-\mathcal{R}_{c}(\mathbf{r})\quad
c=e,h
\label{eq:conteq}
\end{align}
where $\rho_{c}$ and $j_{c}$ are particle and current density respectively,
$\mathcal{G}_{c}$ the generation rate and $\mathcal{R}_{c}$ the recombination
rate of carriers species $c$. The corresponding microscopic conservation law 
reads [\onlinecite{kadanoff:62}]
\begin{align}
&\lim_{2\rightarrow 1}\big\{\left(
i\hbar\left(\partial_{t_{1}}+\partial_{t_{2}}\right)-\left[H_{0}(\mathbf{r}_{1})-H_{0}(\mathbf{r}_{2})\right]
\right)G(\underbar{1}\underbar{2})\big\}\nonumber\\
&=\lim_{2\rightarrow
1}\int_{C}d\underbar{3}\left[\Sigma_{e\gamma}(\underbar{1}\underbar{3})G(\underbar{3}\underbar{2})-
G(\underbar{1}\underbar{3})\Sigma_{e\gamma}(\underbar{3}\underbar{2})\right]
\label{eq:microcont}
\end{align}
In steady state, the continuity equation \eqref{eq:conteq} is reduced to 
\begin{align}
\nabla\cdot
\mathbf{j}_{c}(\mathbf{r})=\mathcal{G}_{c}(\mathbf{r})-\mathcal{R}_{c}(\mathbf{r}).\label{eq:conteqstead}
\end{align}
In the microscopic theory, the divergence of the current corresponds to the
''$<$''-component of the RHS in \eqref{eq:microcont}
\begin{align}
\nabla\cdot
\mathbf{j}(\mathbf{r})
&=-2e\int\frac{dE}{2\pi\hbar}\int d^3 r'\Big[\Sigma_{e\gamma}^{R}
(\mathbf{r},\mathbf{r}';E)G^{<}(\mathbf{r}',\mathbf{r};E)\nonumber\\
&+\Sigma_{e\gamma}^{<}(\mathbf{r},\mathbf{r}';E)G^{A}(\mathbf{r}',\mathbf{r};E)
-G^{R}(\mathbf{r},\mathbf{r}';E)\nonumber\\&\times\Sigma_{e\gamma}^{<}(\mathbf{r}',\mathbf{r};E)-
G^{<}(\mathbf{r},\mathbf{r}';E)\Sigma_{e\gamma}^{A}(\mathbf{r}',\mathbf{r};E)\Big].
\end{align}
If the energy integration is restricted to one carrier species, the above equation corresponds to 
the microscopic version of \eqref{eq:conteqstead}, and expresses the total 
\emph{local} radiative interband transition rate. The total radiative interband
current is found by integrating the divergence over the absorbing/emitting 
volume, and is equivalent to the total \emph{global} transition rate, and, via
the Gauss theorem, to the difference of the current at the boundaries of the absorbing/emitting region. 
On the other hand, the global radiative rate 
equals the total optical rate $R_{\gamma}^{opt}$, related to the global
electromagnetic energy dissipation $w$ through 
\begin{align}
w=-\int_{0}^{\infty}\frac{d\omega}{2\pi}\hbar\omega R^{opt}_{\gamma}(\hbar\omega)
\end{align}
and which, via the quantum statistical average of the Poynting vector operator, can be expressed in terms of  
photon Green's functions and self-energies [\onlinecite{richter:08,
henneberger:09}], providing the following microscopic expressions for absorbed and emitted photon flux
\begin{align}
R_{\gamma}^{opt}(\hbar\omega)=&R_{\gamma,abs}^{opt}(\hbar\omega)-R_{\gamma,em}^{opt}(\hbar\omega),\\
R_{\gamma,abs}^{opt}(\hbar\omega)=&\int d^3 r \int d^3 r'
\left[\Pi_{\gamma}^{
>}(\mathbf{r},\mathbf{r}',\hbar\omega)-\Pi_{\gamma}^{
<}(\mathbf{r},\mathbf{r}',\hbar\omega)\right]\nonumber\\
&\times D_{\gamma}^{<}(\mathbf{r},\mathbf{r}',\hbar\omega),\\
R_{\gamma,em}^{opt}(\hbar\omega)=&\int d^3 r \int d^3 r'
\Pi_{\gamma}^{
<}(\mathbf{r},\mathbf{r}',\hbar\omega)
\hat{D}_{\gamma}^{<}(\mathbf{r},\mathbf{r}',\hbar\omega),
\end{align}
where the photon density of states 
\begin{equation}
i\hat{D}_{\gamma}\equiv i(D_{\gamma}^{>}-D_{\gamma}^{<})
\end{equation}
was introduced. The expression for $R^{opt}_{\gamma,em}$ replaces the
generalized Kirchhoff formula [\onlinecite{wuerfel:82}] for the photon flux
emitted from an excited semiconductor. It allows for the inclusion of a realistic electronic dispersion
via the polarization function $\Pi^{\gamma}$, reflecting the effects of
electron-electron, electron-hole and electron-phonon interactions and the non-equilibrium 
occupation of these states, as well as the consideration of the optical 
modes specific to the geometry of the system.

\section{Implementation}
The general theory outlined above is implemented for quantum well layers inserted in the intrinsic region of 
a $p$-$i$-$n$ diode. Details of this implementation can be found in
[\onlinecite{ae:prb_08}]. As a first approximation, the charge carriers are
treated as a system interacting with a specified equilibrium environment of
photons and phonons, i.e. only the electronic Green's functions are renormalized. 
The main aspects of the model for charge carriers, photons and phonons are given below.

\subsection{Charge carriers}
Electrons and holes are modelled within the framework of empirical tight-binding theory for layered 
semiconductor heterostructures, where the carrier field operators are expanded in planar orbitals 
representing linear combinations of Bloch sums of localized atomic orbitals over the plane of periodicity.
Interactions among carriers are considered only on the Hartree level, equivalent to the solution of the 
macroscopic Poisson equation for the effective potential. Inclusion of screening and excitonic effects would 
require the calculation of two-particle quantities in higher order perturbation
theory. 

Carrier selective contacts are obtained by applying closed system boundary
conditions at minority carrier contacts, which results in a complete supression
of leakage currents.

\subsection{Photons}
Light is described via free field modes with occupation determined by the combination of the spectra of the 
external illumination and the black-body thermal equilibrium with the environment. For practical computations, 
the absorption and stimulated emission  due to the latter can be neglected. The self-energy in the electronic 
Dyson equation describing the coupling to charge carriers is computed within perturbation theory on the level 
of a self-consistent first Born approximation (SCBA) to second order in the
electron-photon interaction.

\subsection{Phonons}
Lattice vibrations are implemented within the harmonic approximation and for noninteracting acoustic as well as 
polar optical modes. The interaction with the electronic system is on the level of a coupling to an equilibrium 
heat bath and is described by the corresponding SCBA self-energy.

\subsection{Absorption, emission and photon flux balance}
The linear absorption coefficient $\alpha$ is related to the transverse
polarization function or photon self-energy via
\begin{align}
\alpha(\hbar\omega)=&-\frac{c}{3\omega\sqrt{\varepsilon_{b}}}\sum_{\mu}\mathrm{Im}
\overleftrightarrow{\Pi}^{\gamma}_{\mu\mu}(\mathbf{q}
=0,\hbar\omega),
\\&\mathrm{Im}\overleftrightarrow{\Pi}^{R}=-\frac{i}{2}(\overleftrightarrow{\Pi}^{>}
-\overleftrightarrow{\Pi}^{<}),
\end{align}
where $\varepsilon_{b}$ is the background dielectric constant. For dipole
coupling to free field photons, the self-energy reads
\begin{align}
\Pi^{\lessgtr}_{\lambda}(\mathbf{q};E)=&i\sum_{\mathbf{k}_{\parallel}}\int\frac{d
E'}{2\pi}\mathrm{Tr}\{\mathbf{M}^{\gamma}(\mathbf{k}_{\parallel},\mathbf{q},\lambda)
\mathbf{G}^{\lessgtr}(\mathbf{k}_{\parallel};E')\nonumber
\\&\times \mathbf{M}^{\gamma}(\mathbf{k}_{\parallel},-\mathbf{q},\lambda)\mathbf{G}^{\gtrless}
(\mathbf{k}_{\parallel};E'-E)\},
\end{align}
where $\mathbf{M}^{\gamma}$ describes the dipolar matter-field coupling and the
trace is over both layer and orbital indices. For the considered model, the total radiative rate is given by
\begin{align}
R_{rad}=\Delta
\sum_{\mathbf{q},\lambda}\Big\{&\phi^{\gamma}_{\lambda,\mathbf{q}}[\Pi^{>}_{\lambda}(\mathbf{q},
\hbar\omega_{\mathbf{q}})-\Pi^{<}_{\lambda}(\mathbf{q},\hbar\omega_{\mathbf{q}})]\nonumber\\
&-\Pi^{<}_{\lambda}(\mathbf{q},\hbar\omega_{\mathbf{q}})\Big\},
\end{align}
where $\Delta$ is the model layer width. From this representation of the rate
follows the intuitive result that the absorbed flux is proportional to the incoming photon 
flux $\phi^{\gamma}$ , while the emitted flux only depends on the polarization
$\Pi^{<}$.

\section{Numerical results} 
 
\subsection{Models system: geometry and simulation parameters}
The simulated system consists of an ultrathin (∼50 nm) $p$-$i$-$n$ diode with a
single quantum well embedded in the $i$-region. A two band tight-binding model is used for the electronic
structure with the the two bulk band gaps given by $E_{g,low}$=0.6 eV and $E_{g,high}$=0.9 eV, with a
symmetric offset of $\Delta E$=0.15 eV. The effective masses and the parameters for the
electron phonon-interaction are chosen for a GaAs/AlGaAs material system. The
system is illuminated at a photon energy of 0.7 eV and an intensity of 1
kW/m$^{2}$.

\subsection{Local density of states, absorption and emission}
\begin{figure}[!t]  
\includegraphics[width=8cm]{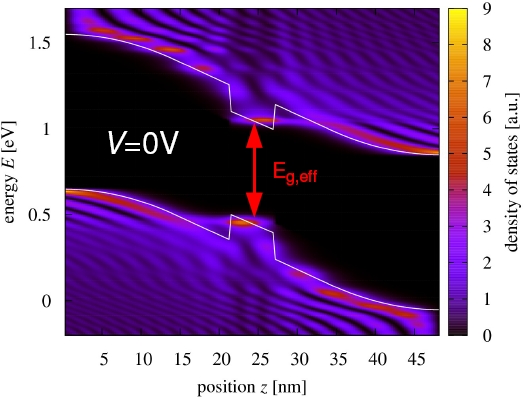}
\caption{Local density of states for single quantum well $p$-$i$-$n$ diode at
vanishing transverse momentum and zero applied bias voltage. The effective band gap in the absorbing region is determined by
 the separation of electron and hole confinement levels.  The interference pattern of the LDOS reflects the closed system boundary
conditions at minority carrier contacts.}
\label{fig:ldos}
\end{figure}
\begin{figure}[!b]  
\includegraphics[width=8cm]{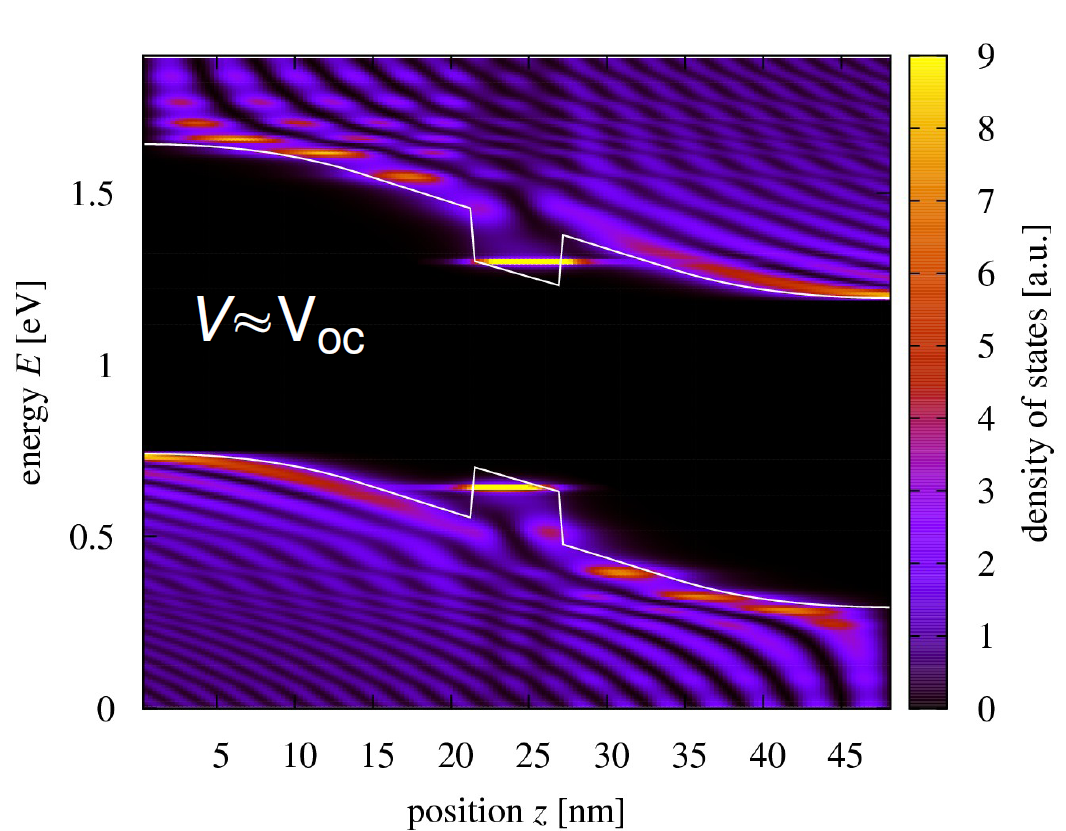}
\caption{Local density of states a large bias voltage close to $V_{OC}$. The
effective band gap exceeds the zero bias value due to a less pronounced Stark effect at the lower effective
field. } 
\label{fig:ldos_voc} 
\end{figure}
\begin{figure}[!t]
\includegraphics[width=8cm]{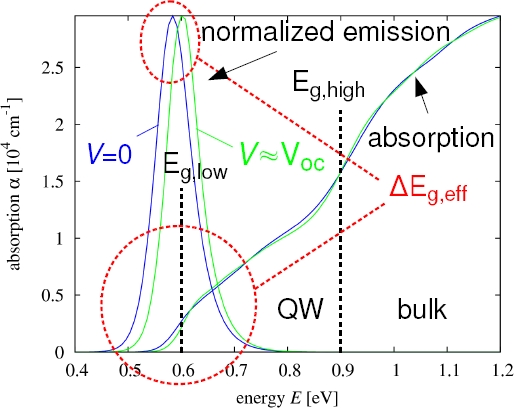}
\caption{The absorption exhibits a significant contribution from the quantum well. There is a pronounced 
blue shift in the fundamental emission edge with increasing voltage due to decreasing field (Franz-Keldysh effect),
with corresponding increase of the effective band gap as observed in the local density of states. }
\label{fig:absem}
\end{figure}

Insertion of a quantum well in the $i$-region of a $p$-$i$-$n$ diode locally
extends the density of states below the high band gap bulk edge, red-shifting the effective absorption edge, which is now given 
by the position of the lowest confinement level, as can be seen in Fig.
\ref{fig:ldos} showing the local density of states at zero bias. Under very high
fields, as in the case considered, the quantum well states hybridize with the bulk continuum outside the well. 
The resulting significant deviation from the
2D-DOS is reflected in the absorption, shown in Fig. \ref{fig:absem}. This so
called Franz-Keldysh effect includes an additional red-shift in the effective band edge, which decreases with the 
effective field at the quantum 
well location under the application of an external bias voltage and resulting in a blue-shift of the emission 
edge as the voltage increases from $V=0$ V to $V=0.44$ V $\approx$ $V_{OC}$,
which is displayed in Fig. \ref{fig:ldos_voc}.

\begin{figure}[!t]
\includegraphics[width=8cm]{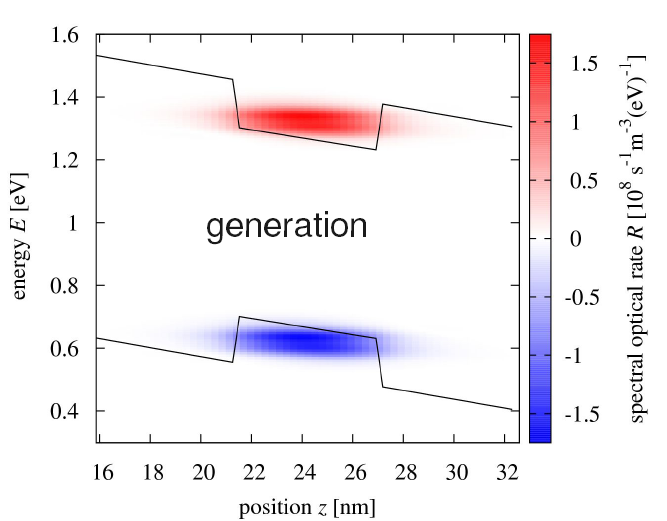}
\includegraphics[width=8cm]{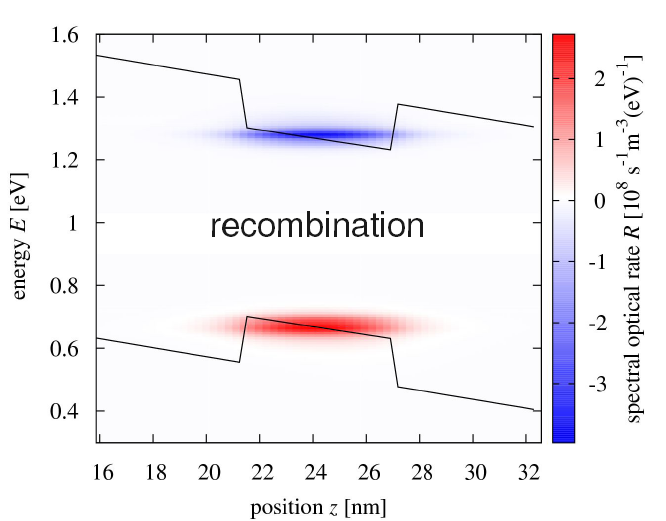}
\includegraphics[width=8cm]{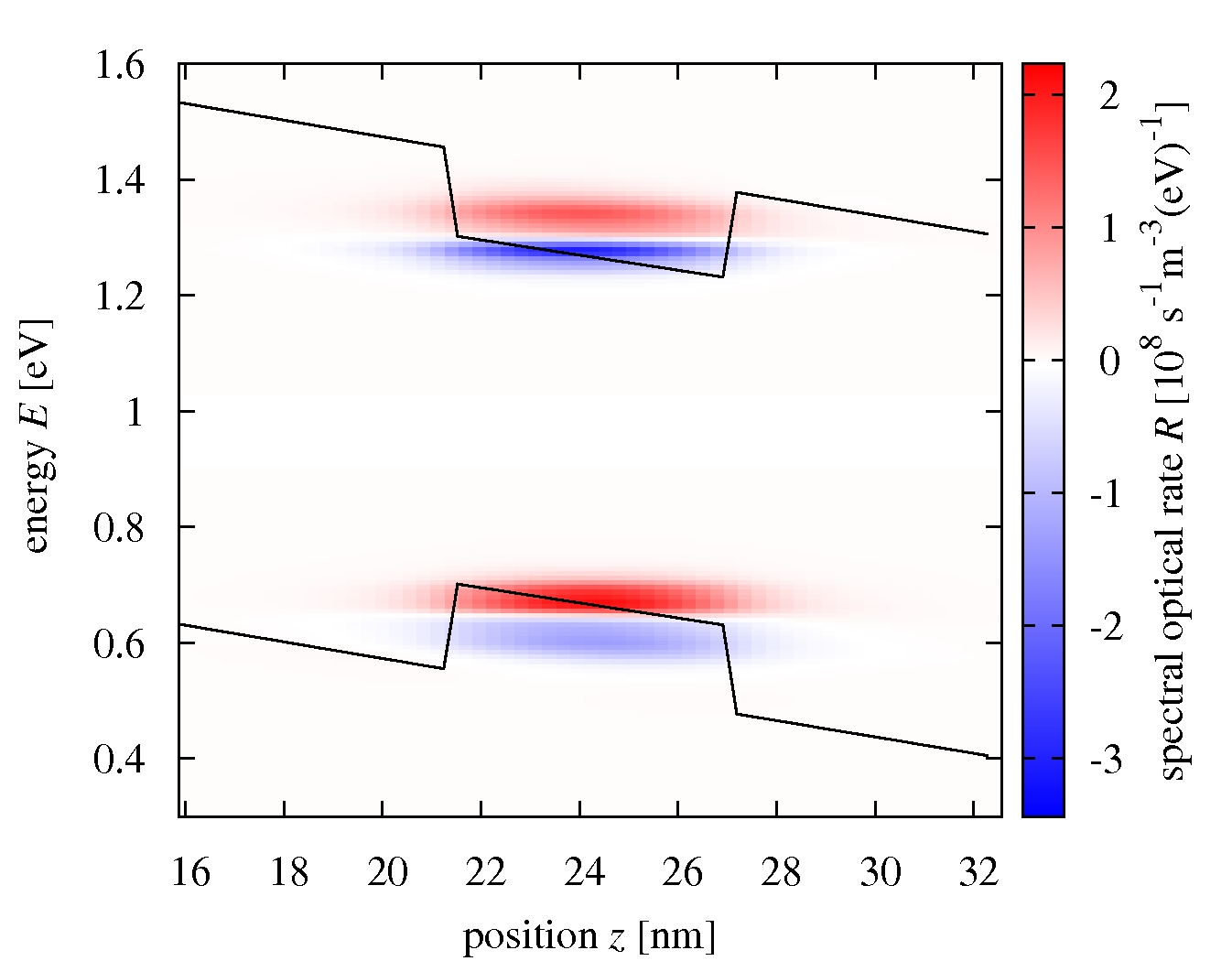}

\caption{Generation, recombination and total radiative interband rate. While
integration over space and energy yields again the terminal current, the spatial extension 
of the electron-photon scattering is completely confined to 
the region of absorbing QW states.}
\label{fig:rates_tot}
\end{figure}

\begin{figure}[!t]
\includegraphics[width=8cm]{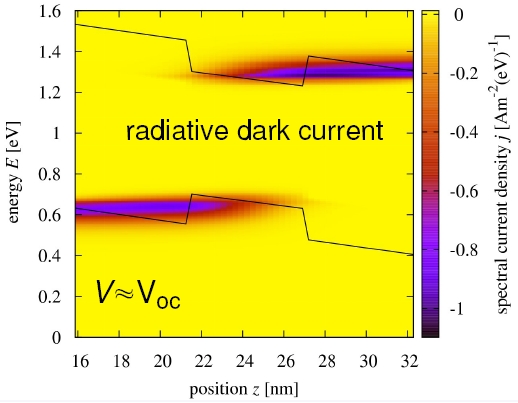} 
\includegraphics[width=8cm]{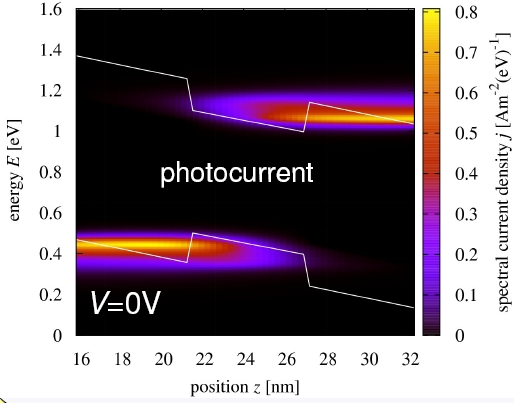}
\includegraphics[width=8cm]{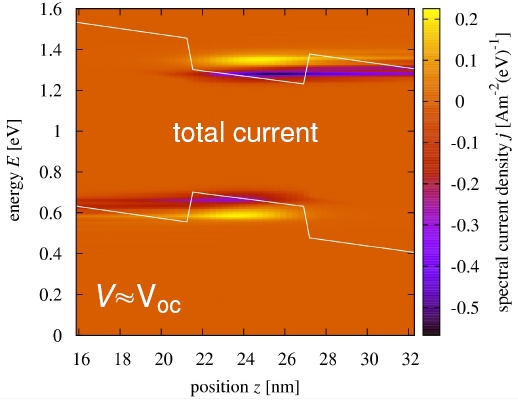}
\caption{The energy resolved local current spectrum in the quantum well region
at short circuit ($V=0$) and close to open circuit conditions ($V=0.44$ V), for
monochromatic illumination of 1 kW/m$^{2}$ with E$_{phot}$=0.7 eV. The total current shows a strong spatial
inhomogeneity with maximum in the well, while the dark and light components 
reach their maximum outside the well.}
\label{fig:current}
\end{figure}
\subsection{Local radiative interband rates}

Fig. \ref{fig:rates_tot} shows the local interband rates for generation,
recombination and the sum of the two effects, as computed via the
electronic Green's functions and self-energies, consisting of inscattering (=generation) and outscattering (=recombination) components.
At illumination photon energies exceeding the effective band gap but below the higher bulk gap, net generation and recombination are 
spectrally well separated and spatially confined to the quantum well, i.e. the region of finite overlap of 
absorbing electronic and hole quantum well states. On the other hand,
integration over energy provides net interband current of the entire device.   

\subsection{Local current spectrum}
The \emph{spectral} difference in generation and recombination rates leads to a
\emph{spatial} inhomogeneity in the generation and recombination currents, as
seen in Fig. \ref{fig:current}, with localization of the maximum in the quantum
well, even though the integrated current is locally conserved, and the separate contributions of radiative dark and photocurrent 
grow towards the contacts and reach their maximum \emph{outside} the absorbing 
region.
  
\section{Conclusions}  
Quantum effects in low dimensional nanostructured absorbers for high efficiency solar cells require new 
approaches for the modelling of absorption and luminescence. A comprehensive microscopic theory of quantum 
photovoltaic devices including the effects of quantum confinement on the optical, electronic and vibrational
properties can be formulated based on the non-equilibrium Green's functions theory of excited semiconductor 
nanostructures. Within this theory, microscopic electronic and optical conservation laws are derived and 
interrelated. Numerical simulations based on an implementation of the theory for quantum well pin-diodes 
provide a picture resolved in space and energy of the generation,
recombination and the corresponding currents in quantum well absorbers at the radiative limit and for both short circuit and close to open 
circuit conditions. 

\bibliographystyle{aiptitle}
\bibliography{/home/aeberurs/Biblio/bib_files/negf,/home/aeberurs/Biblio/bib_files/bandstructure,/home/aeberurs/Biblio/bib_files/generation,/home/aeberurs/Biblio/bib_files/aeberurs,/home/aeberurs/Biblio/bib_files/qwsc,/home/aeberurs/Biblio/bib_files/pv,/home/aeberurs/Biblio/bib_files/scqmoptics}
\end{document}